\documentclass[12pt]{article}
\usepackage{epsf}

\newcommand{\beq}{\begin{equation}}
\newcommand{\eeq}{\end{equation}}
\newcommand{\beqcol}{\begin{array}{rcl}}
\newcommand{\eeqcol}{\end{array}}

\begin{document}

{\thispagestyle{empty}

\begin{flushright}  KCL-MTH-97-68  \end{flushright}
\vfill
\begin{center}
{\Large 
The Form Factors in the Sinh-Gordon Model \\
 \ \                                    }
\\
\vspace{2cm}
{\sc Mathias Pillin}             \\
\vspace{1cm}
Department of Mathematics             \\ 
King's College                     \\
Strand                            \\
London WC2R 2LS, U.K.               \\
\                               \\
e-mail: map@mth.kcl.ac.uk 
\medskip

\vfill
{\bf Abstract}
\end{center}
\begin{quote}
The most general solution to the form factor problem 
in the sinh-Gordon model is presented in an explicit way. 
The linearly independent classes of solutions 
correspond to powers of the elementary 
field. We show how the form factors of exponential operators 
can be obtained from the general solution by adjusting free parameters. 
The general formula obtained in the sinh-Gordon case 
reproduces the form factors of the scaling Lee-Yang model 
in a certain limit of the coupling constant. 
\end{quote}

\eject
}

\setcounter{page}{1}


{\section{Introduction}} 

\bigskip
\bigskip

The famous reconstruction theorem \cite{WIGHT} states that 
once all Wightman functions, i.e. vacuum expectation values 
of local operators, in a physical model are known one can 
reconstruct the entire Hilbert space of the theory and 
the explicit action of the local algebra on it. In a generic 
physical theory the complete knowledge of the Wightman 
functions is almost elusive. 

However, within the last years considerable progress has been 
made in the quantum field theoretical treatment of massive 
integrable models in (1+1) dimensions. In particular there 
are indications that it might be possible, using the 
form factor bootstrap approach \cite{SMIR,KAR-WEI}, 
to compute Wightman functions 
exactly within certain models in the aforementioned class. 

\medskip

Form factors are matrix elements of a local operator ${\cal O}(x)$ 
between multiparticle states and the vacuum. In the case of 
the Sinh-Gordon (sinhG) model, where only one species of particles 
is present, and the multiparticle states can be labeled by the 
rapidities $\theta_i$, the object of study is defined to be

\beq
F_n(\theta_1, \theta_2, \ldots , \theta_n )  \; =  \;
\langle 0 |\;  {\cal O}(0) \; | \theta_1, \theta_2, \ldots , \theta_n 
\rangle .
\label{form-def}
\eeq

It has been shown in \cite{SMIR,KAR-WEI} that due to the 
physical structure of integrable massive models 
in (1+1) dimensions the functions $F_n$ are 
subject to a set of conditions, which are sometimes 
referred to as axioms \cite{SMIR}. These conditions, which 
will be described below in the sinhG case provide equations 
for the matrix elements (\ref{form-def}). 

It has been shown for various models that it is feasible 
to find solutions to these equations; see e.g. [2-17]. 
Due to the quite complicated structure of the form factor 
equations the solutions are sometimes known only up 
to a certain number of incoming particles. 

The form factor equations do not refer to any particular 
local operator in the model. Hence, after having found 
all solutions the next problem is to identify them 
with particular local operators.


The form factors can then at least in principle be used to 
make statements about the operator content of the theory 
and to compute correlation functions of local operators 
in the quantum field theory by:

\beq
\langle 0 | {\cal O}_{1}(x) {\cal O}_{2} (0) | 0  \rangle 
 =  \sum\limits_{n} {\displaystyle {1 \over {n !}}} \int 
F_n^{ {\cal O}_{1}}(\theta_1, \theta_2, \ldots, \theta_n ) 
( F_n^{ {\cal O}_{2}}(\theta_1, \theta_2, \ldots, \theta_n ))^{\ast}  
 e^{ -m |x| \sum_{i=1}^{n} \cosh \theta_i }\prod_{i=1}^{n}
 {\displaystyle{ {\rm{d} \theta_i}\over{2 \pi}}} .
\label{cor-def}
\eeq
 
There is, in fact, some hope that upon a convenient 
parametrization of the form factors the sum in (\ref{cor-def}) 
can be evaluated. This is one of the motivations for the present 
work. For recent progress in this direction see \cite{KOR}.

In this work will give an explicit formula for the most general solution 
to the form factor bootstrap equations in the sinhG model. Our 
method is motivated by \cite{BALOG} where form factors 
of the $O(3)$ nonlinear $\sigma$ model were calculated. 
One might therefore be tempted to think that there exists 
a general method to compute 
form factors of models within the class of two dimensional 
integrable field theories. We are convinced that the formula 
proven in this paper should work with some modifications in 
other affine Toda field theories as well \cite{MAP2}. This 
is due to the fact that the equations to be solved for the 
matrix element (\ref{form-def}) in the sinhG case can be 
reduced to a polynomial recursion equation. This 
polynomial equation is in turn a particular 
specialization of the polynomial recursion equations 
stated in \cite{MAP2} for all ADE affine Toda field 
theories. 

\medskip

Let us comment on the status of the form factors in the 
sinhG model. A few low lying solutions corresponding to 
the elementary field and the trace of the energy momentum 
tensor have been computed in \cite{FMS}. A closed and 
complete solution was found in \cite{KM} for local 
operators satisfying the cluster property. These local 
operators can be identified with exponential fields 
in the theory and in certain limiting cases with 
the elementary field and the trace of the energy momentum 
tensor respectively. However, it turns out that due to 
free parameters which occur in the process of finding solutions 
to the form factor equations there are linearly 
independent solutions which are not covered (at least 
directly) by the solution found in \cite{KM}. In this 
paper we complete the study of form factors in the 
sinhG model in that we find all possible form factor solutions 
in a closed form for the sinhG model. We will show that 
these additional solutions correspond to (renormalized) 
powers of the elementary field.

\medskip

This paper is organized as follows. In section 2 we will review 
some facts about the form factor bootstrap for the sinhG model. 
In section 3 we will prove the main result of this work which is 
the general solution to the polynomial recursion equations which 
follow from the bootstrap equations. For operators satisfying 
the so called cluster property in the sinhG model a solution of 
the form factor equations is known \cite{KM}. We will show 
in section 4 how this particular solution can be obtained from the 
general result by adjusting the free parameters in the 
general solution in a simple way. Since this solution 
corresponds to exponential operators in the sinhG model 
we can identify the linearly independent families 
of the general solutions with renormalized powers 
of the elementary field. This will be done in section 5. 

If the effective coupling in the sinhG model is set to a 
particular value one can build a bridge to the scaling 
Lee-Yang model. In section 6 we will show that even though 
the form factor bootstrap equations are structurally 
different in this case our general result can be used 
to uniquely compute the form factors in this model as well. 
The quantum equivalence of the two possible primary fields 
in this model is then just a consequence of the general 
solution without making any assumptions on the nature of 
the local operators or the structure of the polynomials. 
Since these results have already been obtained using 
different approaches in \cite{ZAM,FMS,KOUB} our presentation 
serves to establish the usefulness of our approach 
to compute form factors in integrable massive models 
in (1+1) dimensions. The last section is left to conclusions.  

\bigskip


\section{The form factor bootstrap in the sinhG model}

\bigskip

In this section some necessary facts about the method to 
calculate form factors in the sinhG model will be recalled. A 
detailed exposition of this topic can be found in \cite{FMS}. 

The sinhG model is the $A_1^{(1)}$ affine Toda field theory 
defined by the relativistically invariant Lagrangian in (1+1) 
dimensions 

\beq 
{\cal L}(x) = \displaystyle{ {1\over{2}} \partial_{\mu} \phi (x) 
\partial^{\mu} \phi (x) - {m^2\over{\beta^2}} \cosh (\beta \phi (x) ) 
} .
\label{lagrange}
\eeq

$m^2$ sets the mass scale of the theory and the coupling $\beta$ is taken 
to be a real number. The bootstrap S-matrix as a function of the 
rapidity $\theta$ is given by \cite{ARIN}

\beq
S(\theta ) = {\displaystyle{ {i \sinh (\theta ) + \sin ( \pi B /2) } 
                  \over {i \sinh (\theta ) - \sin ( \pi B /2) } }} ,
\label{S-matrix}
\eeq

where the ``effective coupling'' is given by $B(\beta) = 
{ {\beta^2 /2 \pi}\over{ 1+ \beta^2 / 4\pi}}$. 

According to the form factor bootstrap method 
of \cite{SMIR,KAR-WEI} the matrix element 
$F_n( \theta_1,\ldots , \theta_n )$, which is of course nothing 
else than a meromorphic function with nontrivial monodromies, 
is subject to the following 
conditions. The first two of them are known as Watson's equations.

\beq
\beqcol 
F_n( \theta_1,\ldots ,\theta_l, \theta_{l+1}, \ldots \theta_n ) 
 & =& S(\theta_l - \theta_{l+1} )  \; 
F_n( \theta_1,\ldots ,\theta_{l+1}, \theta_l , \ldots \theta_n )  \\
& &                   \  \                                        \\
F_n(\theta_1+ 2 \pi i, \theta_2,\ldots , \theta_n ) & = & 
{\displaystyle{\prod^{n}_{l=2}}} S(\theta_l- \theta_1) \times 
F_n(\theta_1,\theta_2,\ldots , \theta_n ) .
\eeqcol
\label{watson}
\eeq

We would like to remark that the second of these equations has 
recently been derived from quantum field theoretical priciples 
\cite{MAX2}.

There is only a single elementary field in the Lagrangian 
(\ref{lagrange}), which has the property to be  self-conjugate. 
From this property one can derive \cite{SMIR,KAR-WEI} the following 
kinematical residue equation  

\beq 
- i\; {\rm res}_{\theta^{\prime} = \theta+ i \pi} \; 
     F_{n+2} (\theta^{\prime}, \theta, \theta_1, \ldots , \theta_n) 
= \left (1 - {\displaystyle{\prod_{l=1}^n}} S(\theta-\theta_l) \right) \;
   F_n(\theta_1, \ldots , \theta_n) . 
\label{kinres}
\eeq

Since there are no fusings in the sinhG model there is -- in contrast 
to other affine Toda field theories \cite{MAP2} -- no bound state 
residue equation. 

In the physical strip we now parametrize the matrix element 
(\ref{form-def}) in full generality, using $x_l = e^{\theta_l}$, 
in the following way. 

\beq
F_n(\theta_1, \ldots, \theta_n ) =H_n \, Q_n(x_1, \ldots, x_n)\; 
{\displaystyle{ 
\prod_{i<j}{ {F^{\rm{min}}(\theta_i-\theta_j) }\over {x_i+ x_j}}}}.
\label{ansatz}
\eeq

$H_n$ is a constant, and the other symbols in this definition 
will be explained in what follows. 

It can now be shown that the procedure of solving the equations 
(\ref{watson}) and (\ref{kinres}) can be split into two steps. From 
(\ref{watson}) we obtain, after requiring 
$Q_n(x_1, \ldots , x_n)$ to be a symmetric function, 
two equations in the minimal form factors $F^{\rm{min}}$ only.

\beq
  F^{\rm{min}} (\theta) = S (\theta) F^{\rm{min}}(-\theta ), \qquad 
  F^{\rm{min}}(2\pi i -\theta) =  F^{\rm{min}}(\theta). 
\label{minform1}
\eeq

This means that we have restored the non-trivial monodromies 
arising from (\ref{watson}) in the minimal form factor. 
These equations can be solved using an integral representation 
of the S-matrix \cite{KAR-WEI,FMS,MAX}. Up to normalization the solution 
is 

\beq
 F^{\rm{min}}(\theta) = {\displaystyle \exp 
  \left( 8 \int_{0}^{\infty} { { {\rm d}t}\over{t} } 
  { { \sinh(t B/2) \sinh( t(2-B)/2 ) \sinh(t)}\over {\sinh^2 (2t) }}  
   \sin^2 ( (i \pi -\theta ) t/\pi ) \right) } .
\label{minform2}
\eeq

This expression can also be expanded into an infinite product of 
$\Gamma$ functions, see \cite{FMS,OOTA}. 

We are now left to solve equation (\ref{kinres}) which, using 
the ansatz (\ref{ansatz}), leads to a polynomial equation. 

\beq
Q_{n+2} ( -x, x; x_1, \ldots, x_n )= (-1)^{n} D_n( x| x_1 ,\ldots , x_n ) 
Q_n( x_1, \ldots , x_n) , 
\label{Q-recurs}
\eeq

where, using the abbreviation $\omega= \exp ( i \pi B /2) $, the recursion coefficient is given by 

\beq
D_n( x | x_1, \ldots , x_n) = {\displaystyle{ 
   { x\over{2 ( \omega - \omega^{-1} ) }} \left( 
  \prod_{l=1}^n ( x + \omega x_l )(x- \omega^{-1} x_l ) - 
  \prod_{l=1}^n ( x - \omega x_l )(x + \omega^{-1} x_l ) \right) }}.
\label{D-def}
\eeq
  
Let us comment on the structure of $Q_n(x_1, \ldots , x_n )$. 
Since (\ref{Q-recurs}) is a purely polynomial recursion equation, 
its solutions $Q_n$ obviously have to be polynomials. Upon 
the requirement mentioned above, $Q_n$ has then to be a symmetric 
polynomial. 

Just by considering (\ref{Q-recurs}) it is not obvious that 
its solutions are symmetric, instead there exists 
a family of non-symmetric ploynomial solutions to this equation. 
But, as it has been outlined, we are not interested in 
these non-symmetric solutions.

\smallskip

As it stands, (\ref{Q-recurs}) does not connect polynomials of even 
$n$ and of odd $n$. The reason for this is the ${\bf Z}_2$ 
symmetry of the sinhG model (\ref{lagrange}). 
However, we will see that the general symmetric solution for 
$Q_{n+2}$ to (\ref{Q-recurs}) to be given in the next section will 
explicitly depend on $Q_{n}$ and also inherit the 
structure of $Q_{n+1}$. The way in which $Q_{n+2}$ 
depends on $Q_{n+1}$ is consistent with the ${\bf Z}_2$ 
symmetry of the sinhG model. 

Obviously (\ref{Q-recurs}) has a one dimensional kernel. 
This means that at each stage of the recursion 
process we will have to take one free parameter 
into account 
We therefore expect the polynomials to be of the following 
structure. 

\beq
\beqcol
Q_{n} &= & A_n Q_{n,n} + A_{n-2} Q_{n,n-2} + \ldots + A_1 Q_{n,1}, 
                  \qquad n \quad {\rm{odd}},  \\
 & &            \ \                           \\
Q_{n} &= & A_n Q_{n,n} + A_{n-2} Q_{n,n-2} + \ldots + A_2 Q_{n,2}, 
                  \qquad n \quad {\rm{even}}.
\eeqcol
\label{Q-structure}
\eeq

\smallskip

The components $Q_{n,l}$ satisfy (\ref{Q-recurs}) independently. 
The coefficients $A_l$ are to be determined by the structure of 
the local operator which is under consideration in (\ref{form-def}). 
This means that the operator content of the quantum field theory 
is linked to these parameters. We will come to this point in 
more detail in sections 4, 5, and 6. 

\medskip

To end this introductory section let us briefly recall some 
facts about symmetric polynomials \cite{MAC} which will be needed 
later. The elementary symmetric polynomials in $n$ variables 
are denoted by $e^{(n)}_l = e^{(n)}_l (x_1, \ldots , x_n )$ 
and defined by the relation $\prod_{r=1}^n (1+ t x_r) = \sum_{l=0}^n 
e^{(n)}_l t^l $. Note that $e_l^{(n)}=0$ for $l < 0$ and for 
$l > n$. If $\lambda= (\lambda_1, \ldots , \lambda_m )$ is a 
partition, we define $E^{(n)}_{\lambda} = e^{(n)}_{\lambda_1} \cdots 
e^{(n)}_{\lambda_m} $. 

The monomial symmetric functions $m^{(n)}_{\lambda} = 
m^{(n)}_{\lambda}(x_1, \ldots , x_n )$ are indexed by 
a partition $\lambda$ and defined by 
$ m^{(n)}_{\lambda} = \sum_{\alpha} x_1^{\alpha_1} \cdots x_n^{\alpha_n}$, 
where the sum runs over all permutations $\alpha$ of the 
partition $\lambda$. 

A special partition which will be needed 
later is $\delta_n = (n-1, n-2, \ldots , 1, 0)$. Taking the 
monomial symmetric functions, we can show that 

\beq
   E^{(n)}_{\delta_n} = m^{(n)}_{\delta_n}  + 
\sum_{\mu < \delta } c_{\mu}  m^{(n)}_{\mu} . 
\label{delta1}
\eeq

The sum goes over partitions $\mu$, satisfying a ``smaller than'' 
relation \cite{MAC} with $\lambda$, $c_{\mu}$ is a 
combinatorial coefficient, which is not important for 
the present purposes. 

As a last remark we would like to add that the kernel polynomial 
of (\ref{Q-recurs}) can be easily written using a Schur function 
$s^{(n)}_{\lambda}=s^{(n)}_{\lambda}(x_1, \ldots , x_n )$. 

\beq
{\displaystyle \prod_{i<j}^{n} (x_i + x_j ) = s^{(n)}_{\delta_n} 
= m_{\delta_n}^{(n)}  + 
 \sum_{\mu < \delta } d_{\mu}  m_{\mu}^{(n)}    
 = E^{(n)}_{\delta_n}  + 
\sum_{\mu < \delta } d^{\prime}_{\mu}  E^{(n)}_{\mu}}. 
\label{schur1}
\eeq

The coefficients $d_{\mu}$ and $d_{\mu}^{\prime}$ are again of 
combinatorial nature \cite{MAC}.

\bigskip


\section{The general solution}

\bigskip

We have shown in the last section that the problem 
of calculating the matrix element (\ref{form-def}) 
reduces to the problem of finding symmetric polynomial 
solutions to the equation (\ref{Q-recurs}). 

\medskip

Let us add two remarks on the structure of the 
solutions $Q_n$. If we require the matrix element 
(\ref{form-def}) to be Lorentz invariant it can be shown 
by standard arguments that the total degree of $Q_n$ is 
$n(n-1)/2$. Its partial degree is $n-1$. However, if 
the operator under consideration is of spin $s$ the 
degree of $Q_n$ is $n(n-1)/2 +s$.

\medskip

We are now going to present the most general solution to 
the recursion equation (\ref{Q-recurs}) in the spinless 
case and comment later on the case of nonvanishing spin.

\medskip

In contrast to other studies, see e.g. \cite{LUK}, 
our approach of calculating the polynomials might seem 
less appealing from the conceptual point of view. However, 
our method of directly calculating the polynomials seems 
to be quite effective and does not suffer from any ambiguities. 
The reader is referred to \cite{MAX3} on this point. 

\medskip

Before presenting the main result of this paper note the 
following convention. In (\ref{Q-structure}) we have been 
indicating the general structure of the polynomials $Q_n$. 
We will denote by $Q_{n}^{\prime}$ the polynomial in 
(\ref{Q-structure}) with all indices of the coefficients 
$A_l$ shifted by $+1$. Consequently $Q_n^{\prime \prime}$ 
will be the polynomial $Q_n$ with all indices at the $A_l$'s 
shifted by $+2$ and so on. For example, if $Q_5 = A_5 Q_{5,5} 
+ A_3 Q_{5,3} + A_1 Q_{5,1}$, then $Q^{\prime}_{5}= A_6 Q_{5,5} 
+ A_4 Q_{5,3} + A_2 Q_{5,1}$. 

Moreover, we indicate by $(x_1, \ldots, \widehat{x_k}, \ldots, 
x_n )$ that the coordinate $x_k$ does not appear inside the 
brackets. 

\medskip

We now state the formula for the general symmetric solution 
of (\ref{Q-recurs}) and will then outline a proof. 

\beq
\beqcol
Q_{n} (x_1,x_2, \ldots , x_n ) & = & {\displaystyle{ 
\sum_{k=2}^{n} D_{n-2} (x_k | x_2,\ldots,\widehat{x_k},\ldots, x_n )
Q_{n-2} (x_2, \ldots ,\widehat{x_k},\ldots, x_n ) }}
{\displaystyle{\prod_{ {l=2}\atop{l \ne k}}^n { {x_1+x_l}\over{x_k - x_l}}}}  
\\ 
& + & {\displaystyle{ \prod_{l=2}^{n} (x_1+x_l) \; \times \; 
Q^{\prime}_{n-1} (x_2, \ldots , x_n ) }} .
\eeqcol
\label{loesung}
\eeq

The first term on the right hand side of this equation is 
motivated by a result obtained in \cite{BALOG} in the 
context of the $O(3)$ nonlinear $\sigma$ model. A term 
of this structure cannot be enough for the sinhG model 
for the following reason. While the total degree 
of the form factor polynomials in the $O(3)$ model is 
the same as in the sinhG model 
for the spinless case, in the former case the partial 
degree is $n-2$. Considering the product factor in the first 
term it is obvious that the highest power of $x_1$ is 
just $n-2$. Therefore something has to be added in the 
sinhG model. We would, however, like to remark that 
for solutions where we set $A_i = 0$, for all $i > 1$, 
the first term is sufficient. This particular solution 
corresponds to the form factors of the elementary field 
according to \cite{FMS}. 

Let us show how to prove the formula. First of all 
it might seem that due to the product in the denomiantor 
in the first term that (\ref{loesung}) is not a polynomial. 
However, if we pull out a factor off the first term, which is just a 
Vandermonde determinant, $a_{\delta}(x_2,\ldots , x_n) = 
\prod_{ {i<j}\atop{i\ge 2}}^{n} (x_i - x_j)$, 
we are left with a polynomial which can 
be shown to be completely antisymmetric in the variables 
$\{ x_2, \ldots , x_n \}$. Hence, this part is divisible 
by  $a_{\delta}(x_2,\ldots , x_n)$ using standard arguments 
\cite{MAC}. This shows that (\ref{loesung}) is actually 
a polynomial which is symmetric in $\{ x_2, \ldots , x_n \}$. 
The second term is by definition a symmetric polynomial in 
these variables. 

\medskip

Obviously, if we set $x_1 = - x_k$ in (\ref{loesung}) 
we will obtain (\ref{Q-recurs}). The first term would 
already be enough to obtain this result, but then, 
as mentioned above, $Q_n$ would in general not be symmetric 
in $x_1$. 

To show that $Q_n$ as defined in (\ref{loesung}) is 
symmetric in $x_1$ as well, it is almost sufficient to prove 
that if we set $x_k = - x_l$ for some $k,l >1$ will 
also lead to the recursion equation (\ref{Q-recurs}). 

To prove this one could start making an ansatz of the 
form (\ref{loesung}) where we replace $Q_{n-1}^{\prime}$ 
by any symmetric polynomial $f^{(n)}(x_2, \ldots , x_n)$ of 
appropriate degree.  

The following identitiy is crucial in the proof. If we 
write 

\beq 
H_{n-2}^{(k)}(x_2, \ldots, x_n) = 
D_{n-2}(x_k | x_2, \ldots, \widehat{x_k}, \ldots , x_n) 
                Q_{n-2} ( x_2, \ldots, \widehat{x_k}, \ldots , x_n), 
\label{H-def}
\eeq

we can show using the properties of the recursion coefficient 
(\ref{D-def}) that 

\beq
 H_{n-2}^{(k)}( x_2, \ldots, x_n) |_{x_l = - x_r} = 
 H_{n-2}^{(r)}( x_2, \ldots, x_n) |_{x_l = - x_k}, \qquad 
l \ne r \ne k .
\label{H-ident1}
\eeq

If we then use the modified ansatz and compute for example 
$Q_{n}(x_1, x_2,x_3, \ldots , x_n )|_{x_2\to -x_3}$ the 
function $f^{(n)}(x_2, x_3, \ldots , x_n)$ has to obey 

\beq
f^{(n)}(-x_3, x_3, x_4, \ldots , x_n) = (-1)^{n-3} 
 D_{n-3} (x_3 | x_4 , \ldots , x_n ) 
f^{(n-2)} (x_4, \ldots , x_n) .
\label{f-ident}
\eeq

Equivalent equations hold if we choose any other two variables 
and take the limit. Obviously (\ref{f-ident}) is nothing other 
than the recursion equation (\ref{Q-recurs}) for $n-1$ variables. 
This shows that $f^{(n)} \propto Q_{n-1}$.  

It can now be shown by applying induction in $n$ to (\ref{loesung}) 
that the constants of the polynomial $Q_{n-1}$ in the expansion 
(\ref{Q-structure}) have to be chosen according to the rule 
mentioned in the beginning of this section 
in order to render the solution polynomial symmetric. 
Note that the term in $Q_{n-1}$ does not merely produce a 
kernel solution to (\ref{Q-recurs}). Our proof shows that 
is it really needed in order to render all linearly 
independent components (\ref{Q-structure}) in $Q_n$ 
symmetric! We will show this explicitly in appendix A.

It is clear that (\ref{loesung}), as it stands,  provides 
due to our analysis of the general 
structure of the solution spaces in 
(\ref{Q-structure}) the most general 
symmetric solution to (\ref{Q-recurs}). This completes 
the proof of relation (\ref{loesung}).

\medskip

Let us make a few remarks on the solution we have found. 

1. Since the degree of $Q_1$ is zero, we set $Q_1(x_1) = A_1$. 
This is the only initial condition we have to choose. 
$Q_2$ already follows uniquely from (\ref{loesung}) upon the 
observation that $D_0 = 0$. We will show in appendix A 
how (\ref{loesung}) works and give explicit expressions 
for the first few polynomials. 

2. It turns out naturally from (\ref{loesung}) that the highest 
component $Q_{n,n}$ (\ref{Q-structure}) of $Q_n$ is always 
the kernel of (\ref{Q-recurs}), i.e. 

\beq 
 Q_{n,n}(x_1, \ldots , x_n) = s^{(n)}_{\delta_n}(x_1, \ldots ,x_n) ,
\eeq

according to (\ref{schur1}).

3. The other components of $Q_n$ admit a description in terms 
of symmetric skew polynomials over the ring 
${\bf Z}[\omega + \omega^{-1}]$; therefore the solution 
polynomials can be thought of as relatives of skew Macdonald 
polynomials \cite{MAC}. It seems that a similar structure is 
present in the polynomial solution spaces of the form 
factors of other affine Toda field theories as well 
\cite{MAP3}, we are, however, not going to elaborate this 
issue here, but rather intend 
to address this interesting mathematical problem 
in a more general context in the future.  

4. It is, of course, possible to give straightfowardly 
an explicit expression for any $Q_n$ just in terms of the $D_r$'s, 
with $r \leq n-2$, $Q_2^{(\prime k)}$, and $Q_{1}^{(\prime l)}$ 
by using (\ref{loesung}) recursively. Since the resulting 
expression is a bit lengthy it will not be written out 
here explicitly. 

5. In order to find a nice expression for any component $Q_{n,l}$ 
in (\ref{loesung}) it is useful to pull out the Vandermonde 
determinant $\prod_{{i<j}\atop{i \ge 1}}^{n} (x_i-x_j)$ 
in (\ref{loesung}). Then one can easily read off $Q_{n,l}$ 
in terms of Schur polynomials, due to the fact that 
a Schur polynomial can be written as \cite{MAC}

\beq 
s^{(n)}_{\lambda} = a_{\lambda+\delta}/a_{\delta}, 
\qquad a_{\lambda+\delta} = \det (x_i^{\lambda_j+n-j} )_{ 1 
\leq i,j \leq n}, \quad a_{\delta} = \prod_{i<j}^{n} (x_i - x_j) .
\label{schur-str}
\eeq

6. According to \cite{CHRISTE,CM,FMS}, solutions for higher spin 
operators can be easily obtained from (\ref{loesung}). Since 
we would like to stress the importance of symmetric skew 
polynomials in the context of form factors we are going to 
write the solution found in \cite{CHRISTE} in terms of just 
one skew Schur polynomial (for details see appendix 
B and \cite{MAC}).

We define the partition $\rho_n= (n^2, n-1, \ldots ,3,2)$. 
Note that for $n=1$ we have a degenerate case, so we 
set $\rho_1=(1)$. The second partition we need is 
$\delta_n$, defined at the end of the previous section. 
Note that both partitions have the property of being 
self-conjugate \cite{MAC}. In order to desribe a 
local operator of spin $2s-1$ we can simply multiply 
the matrix element (\ref{form-def}) with the 
function \cite{CHRISTE,CM,FMS}

\beq
I_n^{2s-1} = (-1)^s \, s^{(n)}_{\rho_s/\delta_s}.
\label{spin}
\eeq

The reason for this is that $I_{n+2}(-x,x,x_1,\ldots , x_n) = 
I_{n}(x_1, \ldots , x_n )$.

\bigskip
\bigskip


{\section{Relation to the cluster property solution}
 
\bigskip

A remarkable solution to the recursive equations (\ref{Q-recurs}) 
has been found in \cite{KM}. We are going to review this 
solution briefly and point out to which local operators it corresponds. 
After that it will be shown how these form factors can be found in 
our general solution (\ref{loesung}). 

\medskip

A local operator ${\cal{O}}(x)$ is said to satisfy the {\it cluster 
property} if, after shifting a certain set of variables by a 
rapidity $\Delta$ and then taking the limit $\Delta \to \infty $, the 
form factors of ${\cal{O}}(x)$ decompose in the following way.

\beq
{\displaystyle{\lim_{\Delta \to \infty}}} F_n^{{\cal{O}}} 
(\theta_1 +\Delta , \ldots , \theta_m + \Delta , \theta_{m+1} , 
\ldots , \theta_n ) \propto 
 F_m^{{\cal{O}}}(\theta_1 , \ldots , \theta_m) 
 F_{n-m}^{\cal{O}}(\theta_{m+1} , \ldots , \theta_n ) .
\label{cluster}
\eeq
 
Note that in general the form factors appearing 
on the right hand side of this equation may correspond 
to local operators different from ${\cal{O}}$.

A family of solutions of (\ref{Q-recurs}) corresponding to 
operators satisfying the property (\ref{cluster}) was found 
in \cite{KM}. 

Introduce the $q$-number symbol 
$[ k]_{\omega} = (\omega^{k} - \omega^{-k} )/ (\omega -\omega^{-1}) $ 
and define the following matrix elements, which depend on a 
parameter $k$ by

\beq
    M_{ij} (k) = e_{2j-i} \, [j-i+k]_{\omega} .
\label{KM-matrix}
\eeq

Solutions of (\ref{Q-recurs}) are then given by the determinant 
of this matrix 

\beq
 Q_n(k) = \det ( M_{ij} (k) ), \qquad i,j= 1, \ldots , n-1 .
\label{KM-loesung}
\eeq

\medskip

As it was mentioned above, the general solution (\ref{loesung}) 
comes naturally in the form (\ref{Q-structure}). In order to 
relate the cluster property solution (\ref{KM-loesung}) 
to (\ref{loesung}) we have to adjust the parameters $A_l$. 

We will be using a combinatorial argument to derive 
the desired relation. First notice that for any $n$ the 
cluster solution (\ref{KM-loesung}) has exactly one term 
of the form $E_{\delta_n}^{(n)}$, as defined in (\ref{delta1}). This 
term has the coefficient $[k]_{\omega}^{n-1}$. Since 
according to (\ref{schur1})  
$\delta_n$ is the leading partition in the expansion of 
the kernel of equation (\ref{Q-recurs}), 
it is clear that the general solution $Q_n$ will have a 
term of the form $E_{\delta_n}^{(n)}$ with coefficient 1 in its 
$A_n$ component. 

We now turn to the recursion coefficient $D_{n}$ which 
was introduced in (\ref{D-def}). We expand $D_n$ in 
powers of $x$ and in terms of monomial symmetric functions. 

In accordance with other affine Toda field theories 
\cite{MAP3} it turns out that in this expansion $D_n$ 
satisfies a kind of duality in the space of partitions. 

This duality is characterized by the fact that in the expansion 
of $D_n$ the coefficients of $x^{k} m^{(n)}_{\lambda}$ and 
$x^{2n -k} m^{(n)}_{\mu}$ with $\lambda + \tilde{\mu}= (2^n)$ 
are identical. In particular it turns out that  the highest 
power of $x$ comes with a symmetric polynomial 
$m^{(n)}_{(1 00 \ldots 0)}$, while the 
lowest power of $x$ has $m^{(n)}_{(222 \ldots 21)}$. According 
to the duality observation, both terms have the same coefficient, 
which is just 1. 

It can now be observed that in the recursion relation 
$Q_n \to D_{n-2} Q_{n-2}$, explicitly stated in (\ref{Q-recurs}), 
the structure of the highest partition in any component 
of $Q_{n-2}$ and the highest and lowest powers of $x$ in the 
expansion of $D_n$ will uniquely induce the highest 
partition in the corresponding component of $Q_n$. 

It can be proven using some simple combinatorics of 
partitions that if $\delta_{n-2}$ is the highest partition 
of a component of $Q_{n-2}$ then the corresponding 
components of $Q_{n}$ will have $\delta_{n}$ as 
its leading partition, apart from $n=5$. 

Moreover, it can be proven that a partition $\delta_n$ 
can never be generated in the recursion process (\ref{Q-recurs}) 
by a partition $\mu$, which satisfies $\mu < \delta_{n-2}$.

\medskip
  
Using these properties it is possible to show inductively which 
components of $Q_n$ in the form (\ref{Q-structure}) do have 
$m^{(n)}_{\delta_n}$, and hence $E_{\delta_n}^{(n)}$, as leading 
term. Note that if $m^{(n)}_{\delta_n}$ or $E_{\delta_n}^{(n)}$ 
occur in any component of $Q_n$ they will come with coefficient 
1 due to the aforementioned structure of $D_n$.  

By direct comparison of the solutions (\ref{KM-loesung}) and 
(\ref{loesung}) we find for the first few coefficients

\beq
A_1 = 1, \quad A_2 = [k]_{\omega}, \quad A_3 = [k]^2_{\omega}, 
\quad A_4 = [k]^3_{\omega} -  [k]_{\omega},\quad A_5 =  [k]^4_{\omega}.
\label{As}
\eeq

Using our combinatorial argument we can now proceed 
inductively and obtain

\beq
  A_{l} = [k]^{l-1}_{\omega} - [k]^{l-3}_{\omega}, \qquad l \ge 6 .
\label{As2}
\eeq

By (\ref{As}) and (\ref{As2}) we have shown that the cluster 
poperty solution (\ref{KM-loesung}) is just a special, but 
physically very important, specialization of our general 
solution (\ref{loesung}). We will elaborate this issue 
in more detail in the next section.

\bigskip
\bigskip


{\section{On the operator content of the sinhG model}}

\bigskip

We are now in a position to identify the local operators 
in the sinhG model with the linearly independent 
solutions of (\ref{loesung}). Again note that (\ref{Q-structure}) 
we encounter one free parameter at each stage of the recursion 
process (\ref{Q-recurs}). Moreover we state that solutions 
$Q_n$ (once given by (\ref{loesung})) for $n$ odd or even 
need to be dealt with seperately due to the ${\bf Z}_2$ 
symmetry. 

From (\ref{As}) and (\ref{As2}) one realizes that there are 
two special values for the parameter $k$. Setting 
$k=0$ will leave $A_1=1$ while $A_l = 0$ for all 
$l \ge 2$. This particular solution corresponds to 
the form factors of the elementary field as has been 
shown by a LSZ analysis in \cite{FMS,KM}. 

Moreover, putting $k=1$ will lead to $A_2=1$ and all 
the other free parameters in the ${\bf Z}_2$ even sector 
vanish. This solution corresponds to the form factors of 
the trace of the energy momentum tensor in the sinhG model 
\cite{FMS,KM}. 

In addition it turns out from (\ref{As}) and (\ref{As2}) 
that for generic values of the effective coupling $B$ 
one cannot project out any other of the linearly independent 
solutions of (\ref{loesung}) by adjusting $k$ in a 
straightforward way. 

\medskip

It was argued in \cite{KM} that the solution $Q_n(k)$ 
given in (\ref{KM-loesung}) corresponds to the 
exponential operators of the form $\exp ( k g \phi (x) )$, 
see also \cite{MS}. We can then take the form factors 
of this exponential operator and expand it in the following 
way (cf. \cite{ACERBI})

\beq
\langle 0 | : e^{k g \phi (0)} : | \theta_1, \ldots , \theta_n 
\rangle = 
{\displaystyle{ 
\sum_{r=0}^{\infty} { {k^r g^r}\over{r ! }} 
\langle 0 | : \phi^r (0) : | \theta_1 , \ldots , \theta_n 
\rangle ,}}
\label{exop}
\eeq

where $: \phi^r (0) :$ indicates a normal ordered (possibly 
renormalized) power of the elementary field in the sinhG model 
(\ref{lagrange}). 

On obvious grounds we require the matrix element 
$\langle 0 | : \phi^r (0) : | \theta_1 , \ldots , \theta_n 
\rangle$ to vanish if $r=0$ or $r > n$. In addition the 
${\bf Z}_2$ invariance of the model dictates that the 
matrix element is nonvanishing only if both $r$ and $n$ 
are either simultaneously even or simultaneously odd. 

\medskip

After adjusting a normalization in order to comply 
with $A_1= 1$ in (\ref{As}), by comparing the powers of 
$k$ in (\ref{exop}) and in (\ref{As}), (\ref{As2}) we find 
the following correspondence 
between the linearly independent solutions characterized 
by the parameter $A_r$ (cf. (\ref{Q-structure})) and the 
local operator $:\phi^r (x) :$ 

\beq
A_r \longleftrightarrow :\phi^r (x) : .
\label{correspondence}
\eeq

Note, however, that this procedure does not fix the 
constants $A_r$ uniquely. 

The correspondence (\ref{correspondence}) was already 
conjectured in \cite{KM}. 
It would be interesting to retrieve this correspondence 
from the UV limit of the sinhG model in the manner it 
was done for the minimal models in \cite{CM,KOUB}.

\bigskip
\bigskip


{\section{Relation to the scaling Lee-Yang model}}

\bigskip

In this section we show that the general formula for 
the polynomial part of the form factors in the sinhG model 
can be applied to the scaling Lee-Yang model (SLY) as well. For details 
on this model we refer to \cite{CM2,ZAM,CM}. It is known that 
this model is integrable, and well studied within the 
context of perturbed conformal field theories. However, it is 
known that this model has some problems with unitarity 
on the quantum field theory level. 

We note that the result established in this section is 
not a new one in the sense that the form factor 
problem for the SLY model was already solved in \cite{ZAM}. 
However, we think it is interesting to outline how 
the general formula (\ref{loesung}) in the sinhG case 
can be applied to the SLY model. Another approach 
to recover the SLY form factors from the sinhG ones 
was given in \cite{FMS}. 

The S-matrix of the scaling Lee-Yang model can be obtained 
from the S-matrix of the sinhG model (\ref{S-matrix}) by specializing 
the effective coupling in the sinhG model \cite{FMS} 
to the value $B =  -2/3$, yielding \cite{CM2}

\beq
{\displaystyle{ S_{SLY} (\theta) = { {\sinh(\theta) + i \sin ( \pi /3)} 
  \over {\sinh(\theta) - i \sin ( \pi /3)}}}} .
\label{SLY-S-matrix}
\eeq
  
Even though the SLY model can be obtained by specialization from 
the sinhG model it has one essentially different feature. As 
in the sinhG case we have just one self-conjugate particle 
of mass $m$. In contrast to the sinhG case the SLY model 
has a nonvanishing three point coupling \cite{CM2}. 
This fact is reflected in the presence of a pole at the rapidity 
$\theta = 2 i \pi /3$ in the S-matrix (\ref{SLY-S-matrix}).

Owing to this additional structure we have one more condition 
on the matrix elements (\ref{form-def}) in addition 
to equations (\ref{watson}) and (\ref{kinres}). We denote the 
vertex of the fusing by $\Gamma = i \sqrt{2} 3^{1/4}$, and 
use the notation $\omega_{SLY} = \exp(i \pi/3 )$. This 
quantity is of course obtained from $\omega$ in the sinhG model 
by using the aforementioned specialization of $B$.
 
The so called bound state residue equation \cite{SMIR} 
which arises due to the presence of a three-point coupling 
is given by

\beq
- i {\rm{res}}_{\theta^{\prime} = \theta} F_{n+1}(\theta^{\prime} + 
i \pi/3, \theta - i \pi/3, \theta_1, \ldots , \theta_{n-1} ) 
= \Gamma \, F_{n}( \theta, \theta_1, \ldots , \theta_{n-1} ) .
\label{bound-res}
\eeq

It was shown in \cite{ZAM} that in the case of the SLY model 
an ansatz similar to (\ref{ansatz}) is possible. 

\beq
F^{SLY}_n ( \theta_1, \ldots , \theta_n ) = H_n Q_{n}^{SLY}(x_1, \ldots , x_n) 
{\displaystyle{ 
\prod_{i < j}^{n} { {f(\theta_i-\theta_j)}\over{x_i + x_j}}}} .
\label{SLY-ansatz}
\eeq

Again we introduce a constant $H_n$ which was given 
explicitly in \cite{ZAM}. As in the sinhG case we can 
determine the degree and the 
partial degree of the polynomials $Q^{SLY}$ by this 
ansatz. It turns out that in the case of spinless 
operators we have that ${\rm{deg}} Q^{SLY}_{n} = n(n-1)/2$, 
while the partial degree of the variables in this polynomial 
is $n-1$. Hence, these criteria are the same as in the sinhG case. 

The functions $f(\theta )$ are the analogues of the minimal 
form factors in the context of the SLY model and 
determined as solutions to (\ref{minform1}). They are 
explicitly given by \cite{ZAM}

\beq
f( \theta ) = { { \cosh(\theta ) -1}\over {\cosh(\theta) + 1/2 }} \,
v(\theta), 
\eeq

with 

\beq
v(\theta) = \exp \left( 4 \int_{0}^{\infty} {{\rm{d}t\over{t}}} 
 { {\sinh (t/2) \sinh(t/3) \sinh (t/6)}\over{\sinh^2(t)}} 
\cosh\left( t + i t \theta/\pi \right) \right).
\eeq

The kinematical residue equation for the SLY model leads to 
an equation for $Q^{SLY}$ which is just (\ref{Q-recurs}) 
with $\omega= \exp( i \pi B/2)$ replaced by $\omega_{SLY}= 
\exp(i \pi/3)$. From now on we assume, unless mentioned explicitly, 
that we are working with $\omega_{SLY}$. 

In contrast to the sinhG model we get one more condition 
on the polynomials $Q^{SLY}$ due to the presence of the 
bound state residue equation (\ref{bound-res}). 

\beq
Q^{SLY}_{n+2} ( \omega_{SLY} x, \omega^{-1}_{SLY} x, x_1, 
\ldots , x_n ) = x {\displaystyle{\prod_{l=1}^{n} (x+ x_l) }} \times 
Q^{SLY}_{n+2} ( x, x_1, \ldots , x_n ) .
\label{Q-bres}
\eeq

The degree of $Q^{SLY}_1$ is zero. In accordance with \cite{ZAM} 
we just set this polynomial equal to one. Using (\ref{Q-bres}) 
we can then easily compute $Q^{SLY}_2$.

\beq
Q^{SLY}_1(x_1) = 1, \qquad    Q^{SLY}_2(x_1,x_2)  = e^{(2)}_1.
\label{Q-1-2}
\eeq

Comparing this result with the general solution in the 
sinhG case it is clear that (\ref{loesung}) will give 
the correct solution for $Q^{SLY}_2$ provided that 
in contrast to using $Q^{\prime}_1$ in the second term on the 
right hand side  we just take the polynomial $Q_1$ itself. 

Since the kinematical residue equations for the sinhG and the 
SLY model are structurally equivalent, it is clear that 
we only have to check which constraints (\ref{Q-bres}) 
puts on the solution (\ref{loesung}) with $\omega$ adjusted to 
the SLY case. 

In order to do that, we have to show under which 
conditions (\ref{loesung}) is consistent with (\ref{Q-bres}), and 
a few identities are necessary. First we notice that 

\beq
D_{n-2} (x_2 | x_3, \ldots,x_{n-2} , \omega_{SLY} x, \omega^{-1}_{SLY} x) 
= (x_2-x)(x_2+x) 
D_{n-3} (x_2 | x_3, \ldots,x_{n-2},  x).   
\label{Dident1}
\eeq

Another useful identity for the recursion coefficient of the kinematical 
residue equation in the SLY case is

\beq
D_{n-2} (\omega_{SLY} x |x_2, x_3, \ldots, x_{n-2}, \omega^{-1}_{SLY} x)
  =  \omega_{SLY} x^3 {\displaystyle{\prod_{l=2}^{n-2} 
           (x+x_l) (\omega_{SLY}^2 x - x_l) }}       
\label{Dident2}
\eeq

It is then tedious but straightforward to show by induction 
that the following formula, which is very similar to (\ref{loesung}), 
generates the unique symmetric polynomial solution to the 
equations (\ref{Q-recurs}) and (\ref{Q-bres}) in the SLY model.

\beq
\beqcol
Q^{SLY}_{n} (x_1,x_2, \ldots , x_n ) & = & {\displaystyle{ 
\sum_{k=2}^{n} D_{n-2} (x_k | x_2,\ldots,\widehat{x_k},\ldots, x_n )
Q_{n-2} (x_2, \ldots ,\widehat{x_k},\ldots, x_n ) }}
{\displaystyle{\prod_{ {l=2}\atop{l \ne k}}^n { {x_1+x_l}\over{x_k - x_l}}}}  
\\ 
& + & {\displaystyle{ \prod_{l=2}^{n} (x_1+x_l) \; \times \; 
Q^{SLY}_{n-1} (x_2, \ldots , x_n ) }} .
\eeqcol
\label{SLY-loesung}
\eeq

Having established this formula, it is sufficient to choose 
only one starting condition, which is as mentioned above 
$Q^{SLY}_1 = 1$. The polynomials for higher particle 
numbers then follow uniquely using (\ref{SLY-loesung}). 
This is analogous to what has been outlined for the sinhG 
case in section 3, but surprising because in the 
SLY case we have two conditions to determine the 
polynomials $Q^{SLY}_n$.

\medskip

It is useful to remark that (\ref{SLY-loesung}) could be 
obtained directly from (\ref{loesung}) by making the 
replacement $\omega \to \omega_{SLY}$ and adjusting 
the parameters $A_l$ in the general form of the 
solutions (\ref{Q-structure}). Arguments similar to 
those used in the previous section allow us to state that 
the specializations

\beq
\omega \to \omega_{SLY}, \quad A_1=A_2=A_3=A_5=1, \quad 
A_4=0, \quad {\rm{and}}  \quad A_l=0, \quad l\ge 6 ,
\label{SLY-repl}
\eeq

will provide the unique (symmetric polynomial) solution 
to (\ref{Q-recurs}) and (\ref{Q-bres}) in the SLY case 
from the solution (\ref{loesung}) in the sinhG case.

The solution polynomials in the SLY case can be given 
in terms of just one skew Schur polynomial (see appendix B 
for details). If we 
want to provide a physical interpretation of this 
result we can split off two elementary symmetric 
polynomials. 

Let us define the following partitions. 

\beq
\beqcol
\lambda^{(n)} & = & ( (n-1)^{n-1}, n-2,n-3, \ldots , 2, 1 ) ,  \\
\rho^{(n)}    & = & ( (n-3)^2, (n-4)^2 ,\ldots , 2^2, 1^2 ),
\eeqcol   \quad
\beqcol
\mu^{(n)} & = &     ( (n-2)^2, (n-3)^2, \ldots , 2^2, 1^2) , \\
\nu^{(n)} & = &     ( (n-4)^2, (n-5)^2, \ldots, 2^2 , 1^2 , 0^2) .
\eeqcol
\label{parti1}
\eeq

We can then write the solution, which was already found 
in \cite{ZAM}, in a compact form

\beq
Q_{n}^{SLY}(x_1, \ldots, x_n) = s^{(n)}_{\lambda^{(n)} / \mu^{(n)}  } 
= e^{(n)}_1 e^{(n)}_{n-1} s^{(n)}_{\rho^{(n)} / \nu^{(n)}  } .
\label{SLY-schur}
\eeq

This is consistent with the result obtained in \cite{ZAM} and 
with an analysis of the form factors of the SLY model in the 
context of cluster property operators in \cite{KOUB}. We would 
like to emphasize that in both approaches certain assumptions 
on the structure of the polynomials $Q_n^{SLY}$ have been 
made. While in the former, a factorization of the polynomial 
solution like the last expression in (\ref{SLY-schur}) 
had been assumed, the latter required the 
local operators in question to be explicitly of cluster type. 
In our approach no assumption was made. We 
have just been using the main result (\ref{loesung}) 
and obtained a unique solution to the form factor 
problem for spinless operators in the SLY model. 

\medskip

Let us comment of the uniqueness of the solution 
(\ref{SLY-schur}). It is believed that 
the form factor approach is suited to the 
classification of the 
operator content of a theory. In general this 
classification is not an easy task. The way to 
do this is to adjust the free parameters, which appear 
when solving the recursion relations for the polynomials 
$Q_n$ to the operator under consideration. As 
in \cite{FMS} one could use LSZ techniques to 
get some information on the factorization properties 
of the polynomials. 

In the case of the SLY model it turns out by 
(\ref{SLY-loesung}) that the recursion relations 
(\ref{Q-recurs}) and (\ref{Q-bres}) do not allow 
for any free parameters. In a standard way of 
thinking one could interpret this result that 
there is only one consistent (non-descendent) local quantum operator 
of zero spin in the SLY model. 

According to a LSZ analysis \cite{FMS,ZAM} the factorization property 
of the solution (\ref{SLY-schur}) indicates 
that the local operator to which our solutions belong should 
be the trace of the energy momentum tensor. The question 
then is, why does the elementary field not appear 
as an independent local operator in our analysis?

The SLY model can be 
understood in the context of perturbed conformal 
models \cite{ZAM,CM2}, from which it is 
clear that the trace of the energy-momentum tensor 
is equal to the elementary field up to a factor. 
Hence, our result gives another confirmation of 
this fact on the quantum field theoretical level. 

\bigskip
\bigskip


{\section{Conclusion and Outlook}}

\bigskip
 
In this paper we have been studying the form factor 
bootstrap of the sinhG model. Having chosen a 
convenient ansatz (\ref{ansatz}) for the matrix elements 
(\ref{form-def}) we arrived at a recursion equation 
(\ref{Q-recurs}) for a polynomial $Q_n(x_1,\ldots , 
x_n)$. The appearance of such polynomial equations 
is a general feature of the form factor bootstrap 
in the case of affine Toda field theories 
\cite{MAP1,MAP2}. By employing a trick \cite{BALOG} 
we have obtained the most general symmetric 
polynomial solution (\ref{loesung}) to these 
recursion equations. From the mathematical 
point of view it turns out that these 
solutions admit a description in terms of 
symmetric skew polynomials over the ring 
${\bf Z}[\omega+\omega^{-1}]$. 

We showed how the remarkable cluster 
property solution \cite{KM} could be obtained by the general 
formula (\ref{loesung}) by adjusting the free parameters 
in a simple way. In turn this gave us an identification 
of the linearly independent solutions of 
(\ref{loesung}) with local operators, which have been 
shown to be the powers of the elementary field. 

In the last section it was pointed out that the general 
formula can as well be applied to the SLY model, yielding 
a unique solution for the polynomials $Q^{SLY}$ with 
no free parameters left. We have given an interpretation 
of this result. 

\medskip

It should be possible to employ the method to compute 
form factors outlined in the present work to find 
solutions to the polynomial recursion equations of 
other affine Toda field theories \cite{MAP2}. However, 
in these cases a major difficulty is the fact that 
more than one species of particles is present and 
that higher order poles do occur. From the viewpoint 
of the theory of symmetric polynynomials \cite{MAC} 
this seems to be an interesting application, since 
from other affine Toda models we will get symmetric polynomials 
in several distinguished classes of variables. 

However, a straightforward application of our method should 
be to treat the Bullough-Dodd model \cite{FMS1,ACERBI}. 
This model is nothing other than the non-minimal version 
of the SLY model and shares the feature with the theories 
which were under consideration in this paper of having 
only one massive field in the Lagrangian.


\bigskip
\bigskip
\bigskip

{\bf Acknowledgement} 

The main result was obtained during a 
stay at the Max-Planck-Institut f\"ur Physik, Munich. I am 
grateful to Profs. J. Wess and W. Zimmermann for the invitation 
and for their kind hospitality. I am indebted to M. Niedermaier 
for many helpful conversations and in particular for 
pointing out the relevance of Ref. \cite{BALOG} to me. 
Thanks go also to G. Watts, M. Flohr, A. Pressley, and R. Sasaki for useful 
discussions. This work was supported by EPSRC (grant GR/L 26216).

\bigskip
\bigskip
\bigskip


{\appendix{{\bf{Appendix A}}}}

\bigskip

In this appendix it is shown how the formula (\ref{loesung}) 
works in practice. We also give explicit results for the 
polynomials up to $Q_5$. 

As it was mentioned in section 3 the polynomial $Q_1$ is 
constant and we have to give a name to this constant

\beq
Q_1(x_1) = A_1 .
\label{q1}
\eeq

The higher solutions will then follow uniquely as will be explained 
now. We take (\ref{loesung}) and use the fact that $D_0 =0$. 
Hence only the second term in (\ref{loesung}) will contribute.

\beq
Q_2 (x_1, x_2 ) = (x_1+x_2) Q^{\prime}_1(x_2) = A_2 e_1^{(2)}
\label{q2}
\eeq

The next step is straightforward, because the second term 
in (\ref{loesung}) will produce merely the kernel solution. 
We have using (\ref{schur1})

\beq
\beqcol
Q_3(x_1,x_2,x_3) & = & D_1(x_2|x_3) Q_1(x_3) {{x_1+x_3}\over{x_2-x_3}} 
  + D_1(x_3|x_2) Q_1(x_2) {{x_1+x_2}\over{x_3-x_2}}   \\
  & + & (x_1+x_2)(x_1+x_3) Q_2^{\prime}(x_2,x_3)      \\
  & = & A_1 e_3^{(3)} + A_3 ( e_2^{(3)} e_1^{(3)} - e_3^{(3)} ) 
    = A_3 s^{(3)}_{(210)} + A_1 e_3^{(3)} .
\label{q3}
\eeqcol
\eeq

In extracting $Q_4$ from (\ref{loesung}) we find that the second 
term will contribute, besides the kernel, a nontrivial term 
which is necessary to supply the terms in $A_2$ arising from 
the first term in (\ref{loesung}) in order to make the 
full $A_2$ component of $Q_4$ symmetric. Using the 
definition (\ref{H-def}) we find

\beq
\beqcol
Q_4(x_1, \ldots, x_4) & = &
       H^{(2)}_2(x_2,x_3,x_4) 
        { {(x_1+x_3)(x_1+x_4)}\over{(x_2-x_3)(x_2-x_4)}} + 
        H^{(3)}_2(x_3,x_2,x_4) 
        { {(x_1+x_2)(x_1+x_4)}\over{(x_3-x_2)(x_3-x_4)}} \\
     & + &     H^{(4)}_2(x_4,x_2,x_3) 
        { {(x_1+x_2)(x_1+x_3)}\over{(x_4-x_2)(x_4-x_3)}}  \\
   & + & (x_1+x_2)(x_1+x_3)(x_1+x_4) 
\left( A_4 s^{(3)}_{(210)}(x_2,x_3,x_4) + A_2 e^{(3)}_3(x_2,x_3,x_4) 
\right)
\eeqcol
\label{q41}
\eeq
      
We can now rewrite the result in terms of symmetric functions 
in four arguments

\beq
Q_4(x_1, \ldots , x_4) = A_4 s^{(4)}_{(3210)} 
   + A_2 e^{(4)}_3 e^{(4)}_2  e^{(4)}_1 .
\label{q42}
\eeq

Computing $Q_5$ goes along similar lines. Here we find that the 
contribution in $A_3$ arising from the first term in (\ref{loesung}) 
is not symmetric, the symmetry is only achieved if we add the 
$A_3$ component of the second term arising from 
$Q_4^{\prime}$. This is a general feature of the recursion 
equation (\ref{Q-recurs}). The kernel solution in $Q_n$ 
will evolve nontrivially, which means that in going from 
$Q_r$ to $Q_s$ with $r<s$ the kernel part of $Q_r$ will 
(apart from $Q_n$ with $n \leq 3$) not be mapped one to 
one onto the kernel of $Q_r$ !

For $Q_5$ we get the following result

\beq
Q_5(x_1, \ldots , x_5 ) = A_5 s^{(5)}_{(43210)} + A_3 
 Q_{5,3} (x_1, \ldots , x_5 ) + A_1 Q_{5,1}  (x_1, \ldots , x_5 ) , 
\label{q51}
\eeq

with 

\beq
\beqcol
Q_{5,3}(x_1,\ldots , x_5) & = & 
e^{(5)}_4 e^{(5)}_3 e^{(5)}_3 
 + e^{(5)}_4 e^{(5)}_4 e^{(5)}_1 e^{(5)}_1 
 + e^{(5)}_5 e^{(5)}_2 e^{(5)}_2 e^{(5)}_1 
 - 2 e^{(5)}_5 e^{(5)}_3 e^{(5)}_2               \\
 & &  - ( 2 + (\omega+\omega^{-1})^2 ) e^{(5)}_5 e^{(5)}_4 e^{(5)}_1 
 + (1+ (\omega+\omega^{-1})^2 ) e^{(5)}_5 e^{(5)}_5, 
\eeqcol
\label{q52}
\eeq

and 

\beq
Q_{5,3}(x_1,\ldots , x_5) =  e^{(5)}_5  e^{(5)}_3  e^{(5)}_2 
         - (\omega+\omega^{-1})^2  e^{(5)}_5  e^{(5)}_5 .
\label{q53}
\eeq

Note that the solution (\ref{q52}) does correspond to 
the local operator $:\phi^3(x):$ and is not covered 
by the results in \cite{FMS}.

According to what has been said in section 3 it is now 
straightforward to give explicit expressions of the 
polynomials $Q_n$ for $n >5$. 

\bigskip
\bigskip

{\appendix{{\bf{Appendix B}}}}

\bigskip

In this appendix we give one (out of many) definition 
of a skew Schur polynomial \cite{MAC}. In doing that 
we do not need to refer to the number of variables 
of the symmetric functions involved. Note that 
skew Schur polynomials arise when one introduces 
a scalar product in the space of symmetric polynomials.

Let $\lambda= (\lambda_1, \lambda_2, \ldots , \lambda_r )$ 
be a partition. The conjugate partition $\lambda^{\prime}$ 
is then a partition obtained from $\lambda$ with entries 
$\lambda_i^{\prime} = {\rm{Card}} \{ j: \lambda_j \geq i \}$. 

Let $\mu$ be another partition. Then the skew Schur 
polynomial can be expressed as a determinant of 
elementary symmetric polynomials in the following way.

\beq
s_{\lambda / \mu} = \det ( 
   e_{\lambda^{\prime}_i - \mu^{\prime}_j -i+j} ).
\eeq

\end{document}